\documentclass[final, conference, twocolumn]{IEEEtran}
\IEEEoverridecommandlockouts

\usepackage{diagbox} 
\usepackage[pdftex]{graphicx}
\usepackage{bm}
\usepackage{cite}
\usepackage{amsmath,amssymb,amsfonts}
\usepackage{mathtools}
\usepackage{cleveref}
\usepackage{autonum}
\usepackage[subrefformat=parens]{subcaption}

\crefname{figure}{Fig.}{Figs.}
\crefname{table}{Table}{Tables}
\crefname{algorithm}{Algorithm}{Algorithms}
\crefname{section}{section}{sections}
\crefname{subsection}{section}{sections}
\crefname{subsubsection}{section}{sections}
\crefname{equation}{}{}

\newcommand{\brak}[1]{\displaystyle \left( {#1} \right)}
\newcommand{\floor}[1]{\displaystyle \left \lfloor {#1} \right \rfloor}

\newif{\ifja}
\jafalse

\newif{\iftd}
\tdfalse
\usepackage{color}
\newcommand{\todo}[1]{\iftd \textcolor{red}{\\(COMMENT: #1)\\} \fi}

\let\%\relax
\DeclareFontFamily{OT1}{pxr}{}
\DeclareFontShape{OT1}{pxr}{m}{n}{<->pxr}{}
\DeclareSymbolFont{letA}{OT1}{pxr}{m}{n}
\DeclareMathSymbol{\%}{0}{letA}{`\%}

\usepackage[noend]{algpseudocode}
\usepackage{algorithm}

\def\BibTeX{{\rm B\kern-.05em{\sc i\kern-.025em b}\kern-.08em
    T\kern-.1667em\lower.7ex\hbox{E}\kern-.125emX}}
\begin{document}

\setlength{\baselineskip}{0.979\baselineskip}
\setlength{\abovedisplayskip}{0.5\abovedisplayskip}
\setlength{\belowdisplayskip}{0.5\belowdisplayskip}

\title{FADEC: FPGA-based Acceleration of Video \\ Depth Estimation by HW/SW Co-design}

\author{
\IEEEauthorblockN{Nobuho Hashimoto\IEEEauthorrefmark{1}, Shinya Takamaeda-Yamazaki\IEEEauthorrefmark{2}}
\IEEEauthorblockA{
\textit{The University of Tokyo}\\
\IEEEauthorrefmark{1}hashimoto-nobuho949@g.ecc.u-tokyo.ac.jp, \IEEEauthorrefmark{2}shinya@is.s.u-tokyo.ac.jp}
}

\maketitle

\begin{abstract}
\ifja
動画像を元に三次元情報を再構築するというタスクは、ロボットやドローンの自律走行のためのナビゲーションや拡張現実 (AR)、3Dモデリングなど幅広いアプリケーションで用いられるため、特に注目されている。このタスクでは、動画像処理特有の処理や DNN (Deep Neural Network) を組み合わせた処理が行われることが多い。近年の深層学習の発展により、精度が向上してきた一方、深層学習の計算回数の多さが原因で、速度の低下や消費電力の増加を引き起こしている。DNN のためには様々なドメイン特化 HW アクセラレータがあるが、動画像処理特有の処理と DNN が交互に実行されるようなアプリケーションの全体の処理を高速化するのは容易ではない。そのため、低消費電力の組み込み環境におけるこのような複雑な処理については FPGA を用いた end-to-end の高速化が求められている。
\else
3D reconstruction from videos has become increasingly popular for various applications, including navigation for autonomous driving of robots and drones, augmented reality (AR), and 3D modeling. This task often combines traditional image/video processing algorithms and deep neural networks (DNNs). Although recent developments in deep learning have improved the accuracy of the task, the large number of calculations involved results in low computation speed and high power consumption. Although there are various domain-specific hardware accelerators for DNNs, it is not easy to accelerate the entire process of applications that alternate between traditional image/video processing algorithms and DNNs. Thus, FPGA-based end-to-end acceleration is required for such complicated applications in low-power embedded environments.
\fi

\ifja
本稿では三次元空間の再構成に用いられる DNN ベースの深度推定手法の一つである DeepVideoMVS を用いた FPGA ベースの新しい高速化手法を提案する。そこで、その手法固有の特性に合わせて、最近の SoC FPGA 上の PL (Programmable Logic) と CPU のような異種のコンポーネントを適切に利用するための HW/SW co-design を用いる。HW 実装するのに適していない演算もあるので、各演算の回数やメモリアクセスのパターンを分析し、HW 実装の容易さと HW によって期待される高速化の度合いという包括的な側面を考慮した上で、SW 実装を行う演算を決める。HW 実装と SW 実装は互いの実行レイテンシーを隠蔽するために、PL と CPU が並列に協調して動作するように実行される。提案したアクセラレータは Xilinx ZCU104 board 上に実装した。その結果、提案手法では、精度の低下を最小限に抑えて、SW のみの実装と比べて 60.2 倍の高速化を達成することができた。
\else
This paper proposes a novel FPGA-based accelerator for DeepVideoMVS, which is a DNN-based depth estimation method for 3D reconstruction. We employ HW/SW co-design to appropriately utilize heterogeneous components in modern SoC FPGAs, such as programmable logic (PL) and CPU, according to the inherent characteristics of the method. As some operations are unsuitable for hardware implementation, we determine the operations to be implemented in software through analyzing the number of times each operation is performed and its memory access pattern, and then considering comprehensive aspects: the ease of hardware implementation and degree of expected acceleration by hardware. The hardware and software implementations are executed in parallel on the PL and CPU to hide their execution latencies. The proposed accelerator was developed on a Xilinx ZCU104 board by using NNgen, an open-source high-level synthesis (HLS) tool. Experiments showed that the proposed accelerator operates 60.2 times faster than the software-only implementation on the same FPGA board with minimal accuracy degradation. Code available: \url{https://github.com/casys-utokyo/fadec/}
\fi
\end{abstract}


\section{Introduction}

\ifja
動画像処理は様々な側面から長年研究されており、古典的な動画像圧縮から、深層学習を用いた物体検知に至るまで広範囲に及んでいる。これらの手法は近年の深層学習の発達により性能が向上し、古典的な手法を上回る精度や品質が実現されている。一方で、動画像を元に三次元情報を再構築するというタスクに関しては、速度・精度の観点からさらに改善の余地があると考えられている。特に、高精度で高密度な三次元情報は、ロボットやドローンの自律走行のためのナビゲーション \cite{drone} や拡張現実 (AR) \cite{ar1,ar2}、3D モデリング \cite{modeling} など幅広いアプリケーションで用いられる重要な情報となっている。しかし、このような情報を取得するためには通常高価なデバイスや複雑な処理が必要となる。これまでよく使用されてきた、LiDAR (Light Detection And Ranging) \cite{lidar} や ToF (Time of Flight) カメラ \cite{tof} などを用いるアクティブセンシングでは、高価なデバイスが必要となり、特定の条件以外ではうまく動作しないといった欠点がある \cite{passive_sensing}。単眼カメラやステレオカメラを用いるパッシブセンシングでは、これらの欠点を克服できる一方、ステレオカメラを用いた場合には複数のカメラを用いた巨大なベースラインや慎重なキャリブレーションが必要となるという欠点もある \cite{passive_sensing}。そのため、一般的な一台の単眼カメラを用いて得られる動画像を元に三次元情報を再構築するというタスクの重要性が増している。
\else
Video processing has been studied for many years, including tasks ranging from classical video compression to object detection. These methods have been improved by the recent development of deep learning, achieving high degrees of accuracy and quality that surpass those of the classical methods. However, for 3D reconstruction from videos, there is room for further improvement in terms of both speed and accuracy. Highly accurate and dense 3D information has become particularly important because it is used in a wide range of applications, such as navigation for autonomous driving of robots and drones \cite{drone}, augmented reality (AR) \cite{ar1,ar2}, and 3D modeling \cite{modeling}. However, obtaining such information usually requires expensive equipment or complex computation. Active sensing, which often uses light detection and ranging (LiDAR) devices \cite{lidar} or time-of-flight (ToF) cameras \cite{tof}, has the disadvantages of requiring expensive devices and specific conditions to work well \cite{passive_sensing}. Passive sensing using monocular or stereo cameras can overcome these disadvantages, but the use of stereo cameras has the disadvantage of requiring large baselines and careful calibrations using multiple cameras \cite{passive_sensing}. Thus, 3D reconstruction from videos obtained using a regular monocular camera is becoming increasingly important.
\fi

\ifja
このような三次元情報の再構築のためには、カメラと対象物の距離を推定するための深度推定が基本となる。DeepVideoMVS \cite{dvmvs} は DNN に基づく、シーンに依存しない深度推定手法であり、つまり、一度学習すればシーンごとに学習をする必要がない。この手法は、動画像処理特有の処理及び RNN (Recurrent Neural Network) の一種である ConvLSTM \cite{convlstm} を含む 96 層の CNN (Convolutional Neural Network) からなる DNN を組み合わせた複合的なアルゴリズムである。DNN のためには様々なドメイン特化 HW アクセラレータがあるが、動画像処理特有の処理と DNN が交互に実行されるようなアプリケーションの全体の処理を高速化するのは容易ではない。そのため、低消費電力の組み込み環境におけるこのような複雑な処理については FPGA を用いた end-to-end の高速化が求められている。
\else
For such 3D reconstruction, depth estimation is the fundamental task for estimating the distance between the camera and target object. DeepVideoMVS \cite{dvmvs} is a DNN-based scene-independent depth estimation method that does not require per-scene training after pre-training. This method is a complex algorithm that combines traditional image/video processing algorithms and DNNs consisting of 96 convolutional neural networks (CNNs), including a recurrent neural network (RNN) called ConvLSTM \cite{convlstm}. Although there are various domain-specific hardware accelerators for DNNs, it is not easy to accelerate the entire process of applications that alternate between traditional image/video processing algorithms and DNNs. Thus, FPGA-based end-to-end acceleration is required for such complicated applications in low-power embedded environments.
\fi
\todo{For such 以降段落を変えて大きく修正}


\ifja
そこで、本稿では DeepVideoMVS を用いた FPGA ベースの新しい高速化手法を提案する。そこで、その手法固有の特性に合わせて、最近の SoC FPGA 上の PL (Programmable Logic) と CPU のような異種のコンポーネントを適切に利用するための HW/SW co-design を用いる。PL と CPU 上の実装は互いに並列に協調して動作するように設計されている。このアクセラレータを Xilinx ZCU104 board 上に実装した。そして、精度の低下を最小限に抑えて、高速に動作することを示した。本稿の主な貢献をまとめると以下の通りである。
\begin{enumerate}
    \item 動画像処理特有の処理及び DNN を組み合わせた DeepVideoMVS という複雑な深度推定手法のための FPGA ベースのアクセラレータを提案した。そのために、最近の SoC FPGA 上の PL と CPU のような異種のコンポーネントを適切に利用するための HW/SW co-design を用いた。
    \item 大規模で実用的なサイズの DNN を含む HW 実装に適した処理について、カスタム回路を高位合成ツールを用いて FPGA 上の PL に設計した。
    \item SW 実装に適した処理について、最適化されたプログラムを CPU 上に設計した上で、HW 実装と SW 実装は互いの実行レイテンシーを隠蔽するために、PL と CPU が並列に協調して動作するように組み合わせた。
    \item 提案手法を Xilinx ZCU104 board に実装した。実行速度は同じ FPGA 上での SW のみの実装より 60.2 倍高速化された。
\end{enumerate}
\else
In this study, we propose a novel FPGA-based accelerator for DeepVideoMVS. We employ HW/SW co-design to appropriately utilize heterogeneous components in modern SoC FPGAs, such as programmable logic (PL) and CPU, according to the inherent characteristics of the method. The implementations on the PL and CPU are designed to work parallelly and cooperatively. We developed this accelerator on the Xilinx ZCU104 board. We also demonstrated that it runs at high speed with minimal accuracy degradation. The main contributions of this study are summarized as follows. \\
\begin{enumerate}
    \item We proposed an FPGA-based accelerator for the complex depth estimation method, DeepVideoMVS, that combines traditional image/video processing algorithms and DNNs. We employed HW/SW co-design to appropriately utilize heterogeneous components in modern SoC FPGAs, such as PL and CPU.
    \item We designed custom circuits for the hardware-friendly processes, including large DNNs with practical sizes, on the PL using a high-level synthesis (HLS) tool.
    \item We designed optimized programs for the software-friendly processes on the CPU and combined the hardware and software implementations for parallel execution on the PL and CPU to hide their execution latencies.
    \item We implemented the proposed accelerator on the Xilinx ZCU104 board. The computation speed was 60.2 times faster than that of the software-only implementation on the same FPGA board.
\end{enumerate}
\fi


\section{Background}
\subsection{Depth Estimation}
\ifja
深度推定に用いられる基本的な原理はステレオマッチングと三角測量である。ステレオマッチングでは視差を推定するが、視差とはある画像の各点とそれに対応するもう一つの画像内の点の位置の差と定義される。次に、三角測量の原理を用いてカメラから各点までの距離を算出する。しかし、対応点を見つけることは容易ではないので、次のように視差を推定するのが一般的である。まず、2 枚の画像のパッチ間の類似度で表されるマッチングコストを算出する。次に、視差の候補を複数用意し、各視差だと仮定した場合のマッチングコストを格納した三次元のコストボリュームを作成する。これを元に画像全体を最適化することで、画素ごとの視差を推定する。
\else
Stereo matching and triangulation are basic principles used for depth estimation. Stereo matching is used to estimate disparities, which are defined as the difference in position between each point in one image and its corresponding point in another image. Triangulation is then used to calculate the distance between the camera and each point. However, because it is not easy to find the corresponding points, it is common to estimate the disparities as follows. First, the matching cost is calculated, which is expressed as the similarity between the patches of the two images. Next, multiple candidate disparities are prepared and a 3D cost volume that stores the matching costs assuming each disparity are created. By optimizing the entire cost volume, the disparity for each pixel is estimated.
\fi
\todo{追加}

\ifja
従来、深度推定には SfM (Structure from Motion) と MVS (Multi-View Stereo) を順番に行う手法が用いられていた。SfM では、カメラ座標からグローバル座標に射影するための $4 \times 4$ の行列として表現されるカメラの pose を推定し、ステレオマッチングの原理を利用して低密度の三次元情報を復元する。その後、MVS で高密度の三次元情報に変換する。しかし、SfM は手ブレやモーションブラー、動体やオクルージョンの影響を受けやすく、MVS でも多くの穴やノイズが発生することが多い \cite{deepv2d,robust_consistent}。そのため、学習型のアルゴリズムも使用されるようになったが、尤もらしい出力を得られるものの、幾何学的整合性に欠けており、必ずしも正確とは言えない。
\else
Traditionally, a method that sequentially performs structure from motion (SfM) and multi-view stereo (MVS) has been used to solve depth estimation tasks. SfM estimates the camera poses, which are represented as a $4 \times 4$ matrix for projection from camera coordinates to global coordinates and recovers low-density 3D information using the stereo matching principle. MVS then transforms this information into high-density 3D information. However, SfM outputs are susceptible to camera shaking, motion blurring, moving objects, and occlusion, and MVS outputs often contain several holes and noises \cite{deepv2d,robust_consistent}. Thus, learning-based algorithms are also used; however, they often produce inaccurate outputs with little geometric consistency, although the outputs are plausible.
\fi
\todo{前半部書き換え}

\ifja
従来の手法と学習型の手法を組み合わせたハイブリッドな手法の研究も行われている \cite{deepv2d,robust_consistent,dynamic_scene_monocular}。これらの手法では、SfM や MVS の処理に DNN を用いることで、より高精度な出力を得ることができる。さらに、以高精度の pose を求めるための最適化を工夫したり、入力や中間出力に pose を用いた幾何学的な補正加えたりすることで、幾何学的な整合性を保てるように工夫している。しかし、これらは計算回数が多い DNN の影響で計算が遅くなるので、DNN の構造や画像の解像度を工夫することで、高速かつ軽量なシステムを目指す研究もある \cite{dvmvs,hitnet}。
\else
Hybrid methods combining traditional and learning-based methods have also been studied \cite{deepv2d,robust_consistent,dynamic_scene_monocular}. These methods achieve more accurate outputs using DNNs for SfM and MVS processes. Furthermore, they can maintain geometric consistency by devising optimization to obtain more accurate poses and adding geometric corrections using the poses to the inputs and intermediate outputs. However, because their computations are slow owing to DNNs that require a large number of calculations, some studies aim for a fast and lightweight system by changing the DNN structure and image resolution \cite{dvmvs,hitnet}.
\fi

\ifja
ここまでのハイブリッド型のアルゴリズムでは、データセットを用いて 1 度学習することで、様々なシーンにおける深度推定を可能にするというものだった。一方で、特定のシーンの静的なまたは動的な三次元空間表現をそのシーンの複数視点の画像や動画などを用いて学習するという深度推定アプローチもある \cite{nerf,nsff,open4d,nr}。しかし、これらの手法では手法の特性上、精度は高くなるが、別のシーンで使用するためには一から学習を行わなければならない。そのため、リアルタイムで様々なシーンに適用するという用途には向かない。
\else
The hybrid algorithms described thus far are designed to enable depth estimation in various scenes by training once with datasets. By contrast, some depth estimation methods learn static or dynamic 3D spatial representations of a particular scene using images or videos from multiple viewpoints \cite{nerf,nsff,open4d,nr}. Owing to the characteristics of these methods, they provide higher accuracy; however, they must be trained again for use in different scenes. Therefore, they are unsuitable for real-time applications in which various scenes are inputted.
\fi

\ifja
この論文では、このような複雑な深度推定の処理を低消費電力の組み込み環境において高速化することを目指している。そのため、シーンごとに学習を行う必要がなく、リアルタイムに近い速度で動作させることが可能だと考えられる DeepVideoMVS を扱うこととする。
\else
This study aims to accelerate such a complex depth estimation algorithm in low-power embedded environments. Thus, we selected DeepVideoMVS, which does not require pre-scene training and is expected to operate at near real-time speed.
\fi

\subsection{DeepVideoMVS}
\begin{figure}[tbp]
    \centering
    \includegraphics[width=\linewidth,pagebox=mediabox]{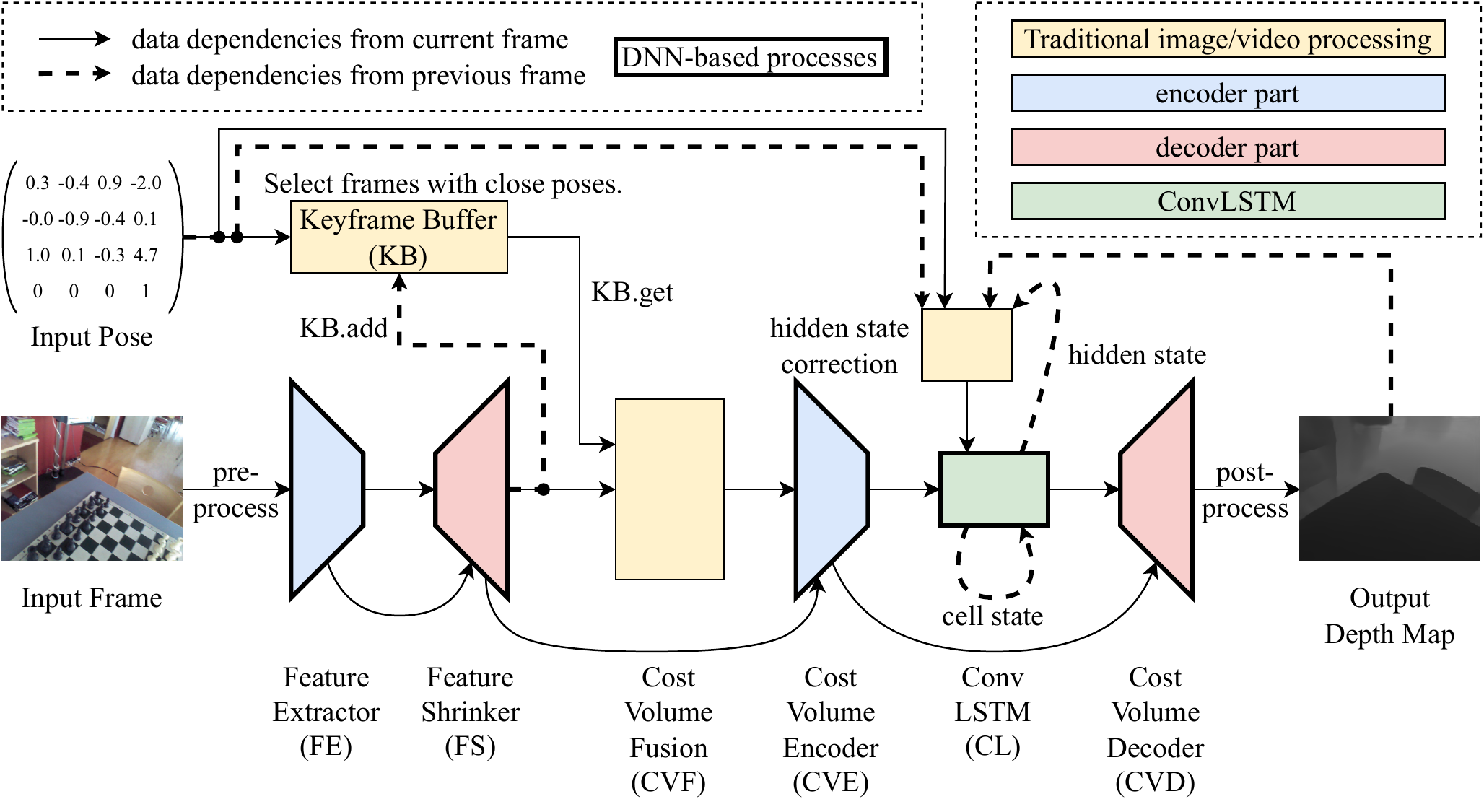}
    \ifja
    \caption{DeepVideoMVS の構成図。元の研究では KB に入力画像を保存していたが、ここでは計算回数を減らすために KB では FS の出力を代わりに格納する。また、矢印はデータの依存関係を表しており、太点線の矢印は一つ前のフレームの処理によって得られた各データからの依存関係を示している。太線で囲まれた領域は主に DNN に基づく処理からなることを表している。}
    \else
    \caption{Diagram of DeepVideoMVS. Although KB stores input images in the original study, KB stores the FS output features instead to reduce the number of calculations in this study. The arrows indicate data dependencies, and the bold dotted arrows show dependencies from each data obtained by processing one previous frame. The areas surrounded by a bold line indicate that the processes are mainly based on DNNs.}
    \fi
    \label{fig:dvmvs}
\end{figure}

\ifja
深度推定の手法として用いた DeepVideoMVS \cite{dvmvs} の構成図を \cref{fig:dvmvs} に示す。この図にはこの手法で行われている処理やデータの依存関係も示されている。この手法では、ビデオとその各ビデオフレームにおける pose を元に、DNN を用いて以前のフレームとステレオマッチングを行うことで深度を推定する。この手法の大きな特徴は、複雑な処理機構の代償として、モデルを学習データセットで一度学習すれば高精度で幅広い状況に適応できることである。ここでは、本稿において必要な処理の概要を説明する (詳細については \cite{dvmvs} を参照されたい)。
\else
\cref{fig:dvmvs} shows a diagram of DeepVideoMVS \cite{dvmvs}, which is used for depth estimation. This figure also shows the processes performed in this method and their data dependencies. This method uses DNNs and estimates depth maps through stereo matching with previous frames using the video and camera poses of each video frame. The key feature of the DNN model is that it can be applied to a wide range of situations with high accuracy once trained using training datasets due to its complex processing mechanism. Here, we outline the processes of the method necessary for this study (see \cite{dvmvs} for further details).
\fi
\todo{2文目追加}

\subsubsection{Processes Based on DNNs}
\ifja
\cref{fig:dvmvs} において、Feature Extractor (FE) は MnasNet \cite{mnasnet}、Feature Shrinker (FS) は FPN (Feature Pyramid Network) \cite{fpn} を利用している。このネットワークでは、画像の特徴量を抽出して receptive field を増やしている。また、Cost Volume Encoder (CVE) と Cost Volume Decoder (CVD) は U-Net \cite{u-net} のようなスキップコネクションを持った encoder-decoder ネットワークのような形をしている。これらは生のコストボリュームを空間方向に正規化するために用いられている。さらに、真ん中に ConvLSTM (CL) を組み込むことで、動画像特有の時系列情報を利用できる設計になっている。CL には非線形活性化関数である sigmoid と ELU \cite{elu} を用いており、BN (Batch Normalization) ではなく layer normalization \cite{ln} を用いている。
\else
In \cref{fig:dvmvs}, the feature extractor (FE) uses MnasNet \cite{mnasnet} and the feature shrinker (FS) uses a feature pyramid network (FPN) \cite{fpn}. These networks extract image features and increase the receptive field. The cost volume encoder (CVE) and cost volume decoder (CVD) are similar to an encoder-decoder network with skipping connections, such as U-Net \cite{u-net}. These are used to normalize the raw cost volume in the spatial direction. Moreover, the incorporation of ConvLSTM (CL) in the middle enables the use of time-series information specific to videos. CL uses the sigmoid and ELU \cite{elu}, which are nonlinear activation functions, and also uses layer normalization \cite{ln} instead of batch normalization (BN).
\fi

\subsubsection{Processes Based on Traditional Image/Video Processing Algorithms}
\ifja
\cref{fig:dvmvs} において、Keyframe Buffer (KB) や Cost Volume Fusion (CVF)、hidden state の補正は主に動画像特有の処理に関わっている。KB には画像の特徴量である FS の出力がその時のカメラの pose 情報と共に保存されている。そして、その特徴量を pose が近いフレームが入力された時に取り出して、再利用する。この処理により、以前見た同じような場面の情報が利用できるため、精度の向上に役立つ。CVF では、まず、過去と現在の pose を元に、grid sampling を行うことで、過去の特徴量を現在の視点から見たものに変換する。その結果を利用して、現在の特徴量から cost volume を求める。同様に、以前の hidden state に視点の変化を反映させるためにも grid sampling を行っている。この grid sampling は以下のように表される。
\begin{align}
    (i, j) =& (\floor{\bm{g}_{s,t,0}}, \floor{\bm{g}_{s,t,1}}) \\
    (k, l) =& (\bm{g}_{s,t,0} - \floor{\bm{g}_{s,t,0}}, \bm{g}_{s,t,1} - \floor{\bm{g}_{s,t,1}}) \\
    \bm{y}_{s,t} =& (1-k) \cdot (1-l) \cdot \bm{x}_{i,j} + (1-k) \cdot l \cdot \bm{x}_{i,j+1} \\
    &+ k \cdot (1-l) \cdot \bm{x}_{i+1,j} + k \cdot l \cdot \bm{x}_{i+1,j+1}
\end{align}
ここで、$\bm{g}$ は grid、$\bm{x}, \bm{y}$ は入出力、$(i, j), (k, l)$ は $\bm{g}$ の整数部分、小数部分をそれぞれ表している。
\else
In \cref{fig:dvmvs}, the keyframe buffer (KB), cost volume fusion (CVF), and hidden state correction are mainly related to traditional image/video processing algorithms. KB stores the FS output, which is an image feature, along with the camera pose. The feature is retrieved and reused when a frame with a similar pose appears as input. This process can help improve the accuracy because information on previous similar scenes is available. CVF first converts past features into those observed from the current viewpoint by grid sampling using past and current poses. The converted features are used to obtain the cost volume from the current feature. Similarly, grid sampling is also performed to apply viewpoint changes to the previous hidden state. Grid sampling is formulated as follows:
\begin{align}
    (i, j) =& (\floor{\bm{g}_{s,t,0}}, \floor{\bm{g}_{s,t,1}}) \\
    (k, l) =& (\bm{g}_{s,t,0} - \floor{\bm{g}_{s,t,0}}, \bm{g}_{s,t,1} - \floor{\bm{g}_{s,t,1}}) \\
    \bm{y}_{s,t} =& (1-k) \cdot (1-l) \cdot \bm{x}_{i,j} + (1-k) \cdot l \cdot \bm{x}_{i,j+1} \\
    &+ k \cdot (1-l) \cdot \bm{x}_{i+1,j} + k \cdot l \cdot \bm{x}_{i+1,j+1},
\end{align}
where $\bm{g}$ denotes a grid, $\bm{x}$ and $\bm{y}$ denote the input and output, respectively, and $(i, j)$ and $(k, l)$ denote the integer and fractional parts of $\bm{g}_{s,t}$, respectively.
\fi

\subsubsection{Dataflow}
\ifja
\cref{fig:dvmvs} において、入力の pose、CL の cell state と hidden state、及び最終出力の depth map は、次の入力フレームの処理に引き継がないといけない。KB においては、CVF の入力を得るためには入力の pose のみで十分だが、次の入力画像を処理するまでには、FS の出力を格納しておく必要がある。このように、各処理がどのタイミングまでに終われば良いかを考える。入力が連続的に与えられる場合には特に、データフローを深く理解することが全体の処理を高速化するために重要となる。
\else
In \cref{fig:dvmvs}, the input pose, CL cell state, CL hidden state, and final output depth map must be passed on to the processing of the next frame. In KB, the input pose alone is sufficient to obtain the input for CVF; however, the FS output must also be stored before the next input image is processed. In this manner, we consider the timing of when each process needs to be completed. A deep understanding of the dataflow is essential for accelerating the entire process, particularly when inputs are given consecutively.
\fi
\todo{大幅に修正}

\section{FADEC}
\ifja
本稿では、低消費電力の組み込み環境における、DeepVideoMVS の FPGA を用いた end-to-end の高速化手法を提案する。最近の SoC FPGA には PL と CPU が載っている。PL は PYNQ のようなフレームワークを使用して簡単に操作でき、CPU コアは高速に動作し、これらはメモリを共有している。そのため、近年 HW/SW co-design は一般的になってきた。そこで、全体の処理の高速化を行うためには、適切なタイミングで PL と CPU に各処理を割り当てることが重要である。
\else
This study proposes FPGA-based end-to-end acceleration for DeepVideoMVS in low-power embedded environments. Modern SoC FPGAs contain PL and CPU; the PL can be easily manipulated using a framework such as PYNQ, the CPU core is fast, and they share memory. Thus, HW/SW co-design has become common in recent years. To accelerate the entire process, it is important to assign each process to the PL and CPU at an appropriate time.
\fi

\ifja
全体の手順は以下の通りである。1) HW/SW Co-design: それぞれの演算を HW か SW かどちらで実装すべきかを、HW と SW のそれぞれの特性を活かして検討する。2) HW Design: HW 実装に適した処理について、カスタム回路を高位合成ツールを用いて FPGA 上の PL に設計する。3) SW Design: SW 実装に適した処理について、最適化されたプログラムを CPU 上に設計する。4) HW/SW Scheduling: HW 実装と SW 実装は互いの実行レイテンシーを隠蔽するために、PL と CPU が並列に協調して動作するように組み合わせる。
\else
The overall procedure is as follows: 1) HW/SW Co-design: We examine whether each operation should be implemented in hardware or software, taking advantage of the characteristics of hardware and software. 2) HW Design: We design custom circuits for the hardware-friendly processes on the PL using an HLS tool. 3) SW Design: We design optimized programs for the software-friendly processes on the CPU. 4) HW/SW Scheduling: We combine the hardware and software implementations such that they are executed in parallel on the PL and CPU to hide their execution latencies.
\fi
\todo{1段落目を追加、2段落目を簡略化}
\todo{日本語直す}

\subsection{HW/SW Co-design}
\ifja
ここでは、まず各演算の回数やメモリアクセスのパターンを分析し、そして HW 実装の容易さと HW によって期待される高速化の度合いという包括的な側面を考慮する。最後に、それぞれの演算を HW か SW かどちらで実装すべきかについて、HW と SW のそれぞれの特性を活かして検討する。
\else
We first analyze the number of times each operation is performed and its memory access pattern, and then consider comprehensive aspects: the ease of hardware implementation and degree of expected acceleration by hardware. Finally, we examine whether each operation should be implemented in hardware or software by taking advantage of the characteristics of hardware and software.
\fi

\subsubsection{Number of Multiplications}
\begin{table}[tbp]
    \ifja
    \caption{DeepVideoMVS において主要な各処理で行われている演算の回数。conv の右の数字は (kernel size, stride) を表している。}
    \else
    \caption{Number of operations performed in each major process of DeepVideoMVS. The numbers to the right of \textit{conv} represent (kernel size, stride).}
    \fi
    \label{tb:operations}
    \centering
    \begin{tabular}{c|rrrrrr}
        \hline \hline
        \diagbox{Operation}{Process} &  FE  &        FS  &        CVF  &        CVE  &       CL &        CVD  \\
        \hline
        Conv (1, 1)           & \textbf{33} & \textbf{5} &           0  &          0  &         0  &          0  \\
        Conv (3, 1)           &  \textbf{6} & \textbf{4} &           0  &  \textbf{9} & \textbf{1} & \textbf{14} \\
        Conv (3, 2)           &  \textbf{2} &         0  &           0  &  \textbf{3} &         0  &          0  \\
        Conv (5, 1)           &  \textbf{7} &         0  &           0  &  \textbf{3} &         0  &  \textbf{5} \\
        Conv (5, 2)           &  \textbf{3} &         0  &           0  &  \textbf{1} &         0  &          0  \\
        Activation (ReLU)     & \textbf{34} &         0  &           0  & \textbf{16} &         0  & \textbf{14} \\
        Activation (sigmoid)  &          0  &         0  &           0  &          0  & \textbf{3} &  \textbf{5} \\
        Activation (ELU)      &          0  &         0  &           0  &          0  & \textbf{2} &          0  \\
        Addition              & \textbf{10} & \textbf{4} & \textbf{128} &          0  & \textbf{1} &          0  \\
        Multiplication        &          0  &         0  &  \textbf{64} &          0  & \textbf{3} &          0  \\
        Concatenation         &          0  &         0  &           0  &  \textbf{4} & \textbf{1} &  \textbf{5} \\
        Slice                 &          0  &         0  &           0  &          0  & \textbf{4} &          0  \\
        Layer Normalization   &          0  &         0  &           0  &          0  & \textbf{2} &  \textbf{9} \\
        Upsampling (nearest)  &          0  & \textbf{4} &           0  &          0  &         0  &          0  \\
        Upsampling (bilinear) &          0  &         0  &           0  &          0  &         0  &  \textbf{9} \\
        Grid Sampling         &          0  &         0  & \textbf{128} &          0  &         0  &          0  \\
        \hline
    \end{tabular}
\end{table}

\begin{figure}[tbp]
    \centering
    \includegraphics[width=0.8\linewidth,pagebox=mediabox]{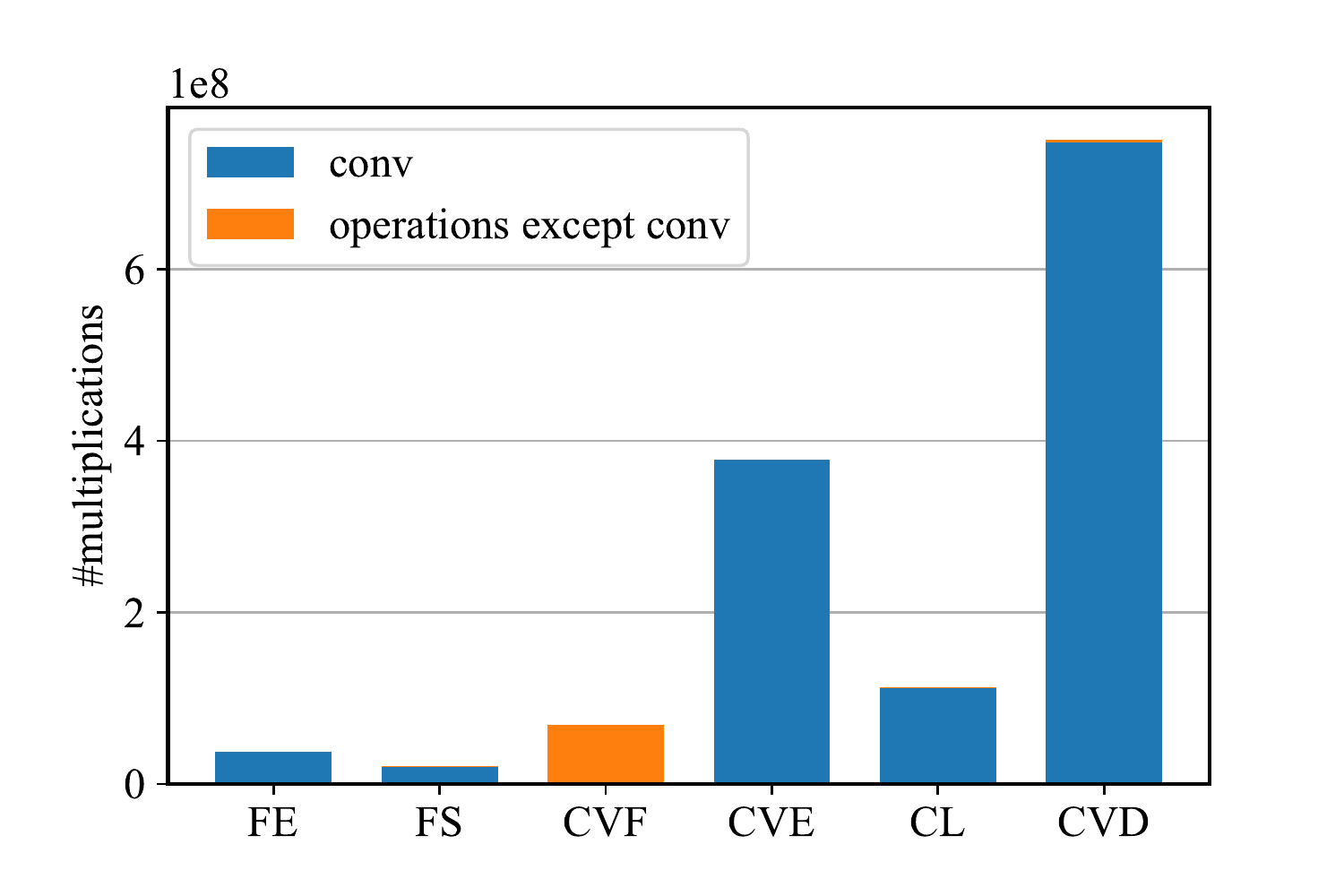}
    \ifja
    \caption{DeepVideoMVS において主要な処理の乗算の回数についての比較。}
    \else
    \caption{Comparison of the number of multiplications performed in each major process of DeepVideoMVS.}
    \fi
    \label{fig:muls}
\end{figure}

\ifja
DeepVideoMVS において、主要な各処理で行われている演算をまとめると \cref{tb:operations} のようになる。これらは主に加算と乗算からなる。そこで \cref{fig:muls} では、テンソルのサイズを考慮した上で、加算よりもボトルネックになる乗算の回数に関してまとめている。コストボリュームの処理である CVE 及び CVD を合わせると、全体の乗算回数のうち 82.4$\%$ を占めていることがわかる。KB へのデータの格納や hidden state の補正のようなその他の処理についてはほとんど計算を行う必要がないので、記載していない。これらを踏まえると、特に CVE 及び CVD を優先的に HW で高速化を行うことがボトルネックの解消に役立つと考えられる。これらの処理において、conv の計算に必要な乗算が全体の乗算の 99$\%$ 以上を占めている。そのため、conv に関しては HW での高速化が必須であり、これによって他の処理も高速化される。従って、conv の計算がない CVF がボトルネックになる。
\else
\cref{tb:operations} outlines the operations performed in each major process of DeepVideoMVS. These operations mainly consist of addition and multiplication operations. Thus, \cref{fig:muls} summarizes the number of multiplications, which are more of a bottleneck than an addition, considering each tensor size. CVE and CVD for cost volume processing together make up 82.4$\%$ of the total multiplications. Other processes, such as storing data in KB and correcting hidden states, are not included because they require few calculations. Thus, CVE and CVD should be prioritized for hardware acceleration to eliminate bottlenecks. In these processes, the multiplications required for the \textit{conv} calculation account for more than 99$\%$ of the total multiplications. Thus, it is essential to accelerate \textit{conv} in hardware, which would also accelerate the other processes. Eventually, CVF, which does not involve \textit{conv}, becomes a bottleneck.
\fi

\subsubsection{Memory Access Pattern}

\ifja
次に、\cref{tb:operations} にある演算のメモリのアクセスパターンについて考える。
\textit{Conv and upsampling}: sliding window の要領でアクセスが行われる。このアクセスパターンはデータの再利用性が高く、特に conv は演算律速であるため、HW 実装に適している。\\
\textit{Activation}: 通常 conv に fold されるので、アクセスパターンは考えなくてよい。 \\
\textit{Addition and multiplication}: element-wise operator なので、メモリバンド幅律速である。そのため、HW でも SW でも性能に大きな違いはない。\\
\textit{Concatenation and slice}: ほぼ連続的にメモリアクセスを行うので、メモリバンド幅律速である。そのため、HW でも SW でも性能に大きな違いはない。\\
\textit{Layer normalization}: 全ての値を走査して各層の平均と分散を求め、再度全ての値を走査して正規化を行う。つまり、各要素が 2 回ずつアクセスされるため、これまでの演算よりメモリバンド幅 intensive になる。こちらも HW でも SW でも性能に大きな違いはない。\\
\textit{Grid sampling}: element ごとに指定された座標の値を bilinear interpolation をして取得するという処理を行う。そのため、どの座標が指定されるかによってアクセスパターンが大きく変わってしまう。このアクセスはランダムに近いものになるため、HW を用いて高速化することは難しい。
\else
Next, we examine the memory access patterns of the operators in \cref{tb:operations}. \\
\textit{Conv and upsampling}: These operators access memory in the manner of a sliding window. This access pattern has high data reuse, and \textit{conv} is particularly suitable for hardware implementation because of the bottleneck in its operations. \\
\textit{Activation}: As this is usually folded into \textit{conv}, it is unnecessary to consider its memory access. \\
\textit{Addition and multiplication}: As these are element-wise operators, their bottlenecks are in memory bandwidth. Thus, there is no significant difference in the performance between the hardware and software executions. \\
\textit{Concatenation and slice}: As these memory access patterns are almost sequential, their bottlenecks are in memory bandwidth. Thus, there is no significant difference in the performance between the hardware and software executions. \\
\textit{Layer normalization}: All values are scanned to obtain the mean and variance of each layer; this scanning is then repeated for normalization. Thus, each element is accessed twice, which makes this operator more intensive in memory bandwidth than previous ones. There is also no significant difference in the performance between the hardware and software executions.  \\
\textit{Grid sampling}: This operator performs bilinear interpolation to obtain the value at the position represented by the coordinates specified for each element. Thus, the access pattern varies significantly, depending on the specified coordinates. As this operator requires irregular access, accelerating it by hardware is difficult.
\fi

\subsubsection{HW/SW Partitioning}
\ifja
それぞれの処理を HW か SW かどちらで実装すべきか決定する。\cref{tb:operations} の演算のうち、以下の 6 つの演算 conv, activation (sigmoid), activation (ELU), layer normalization, upsampling (bilinear), grid sampling について HW 実装を検証する。conv は量子化と並列化を行うことで HW で高速に動作させることができる。非線形活性化関数 (sigmoid, ELU) については、LUT を用いて exponential の計算を近似することで、軽量に実装を行うことができる。一方で、layer normalization については、各層を正規化するために平方根演算や割り算を必要とするため、高精度を維持しながら HW で高速化することは難しい。upsampling (bilinear) についても、メモリアクセスパターンが若干規則的ではなく、高精度を維持するためには浮動小数点演算が望ましいため、HW で大幅に高速化するとは考えづらい。そのため、精度を担保するためにも SW で浮動小数演算を用いた実装を行っても良いと考えられる。最後に、grid sampling については bilinear interpolation も必要で、かつメモリアクセスもほぼランダムなので、SW で実装を行う方が望ましいと考えられる。
\else
We determine whether each process should be implemented in hardware or software. Among the operations in \cref{tb:operations}, we examine the hardware implementation of the following six operations: \textit{conv}, \textit{activation (sigmoid)}, \textit{activation (ELU)}, \textit{layer normalization}, \textit{upsampling (bilinear)}, and \textit{grid sampling}. Hardware acceleration methods for \textit{conv} through quantization and parallelization are well known. The nonlinear activation functions (\textit{sigmoid} and \textit{ELU}) can be implemented in a lightweight manner by approximating the exponential functions using LUTs. By contrast, \textit{layer normalization} is difficult to accelerate using hardware while maintaining high precision because it requires square root and division operations to normalize each layer. It is also unlikely that \textit{upsampling (bilinear)} can be significantly accelerated by hardware because the memory access pattern is slightly irregular and floating-point arithmetic operations are suitable for high precision. Thus, it is considered acceptable to implement it in software by using floating-point arithmetic to ensure precision. Finally, because \textit{grid sampling} requires bilinear interpolation and its memory access pattern is irregular, the software implementation is preferable for this operation.
\fi

\ifja
各処理を HW と SW に分類した結果は以下のようになる。\\
\textit{FE、FS、CVE、及び CVD}: 主に conv からなるため、upsampling (bilinear) を除いて HW で実装することが望ましい。\\
\textit{CL}: layer normalization を除いて HW で実装することが望ましい。\\
\textit{CVF}: この処理ではまず 64 回の grid sampling を 2 回行い、それらの結果を足し合わせて、現在の特徴量と同じ形の 64 個のテンソルを得る。それらのテンソルをそれぞれ現在の特徴量と掛け合わせて、チャネル方向に足し合わせることで、cost volume を得る。grid sampling は SW で実行されるため、現在の特徴量を SW から HW に送り、残り処理を SW で行い、その結果を HW に渡すのは合理的である。これにより、残りの処理を HW で行った場合に比べて、浮動小数演算を使用することにより精度を担保した上で、HW・SW 間の通信を 2/64 に削減することができる。さらに、CVF の乗算回数が全体に対して占める割合も 5.0$\%$ なので、多少計算が遅くても HW の処理とできるだけ並列化することで、実行レイテンシーを隠蔽できると考えられる。\\
\textit{\cref{fig:muls} に含まれない演算}: 必要な計算回数が少ないため、SW で実装することで実装を簡略化する。
\else
The HW/SW partitioning of the processes in \cref{fig:muls} is summarized as follows: \\
\textit{FE, FS, CVE, and CVD}: As these are mainly composed of \textit{conv}, these should be implemented in hardware except for \textit{upsampling (bilinear)}. \\
\textit{CL}: This process should also be implemented in hardware except for \textit{layer normalization}. \\
\textit{CVF}: In this process, 64 \textit{grid sampling} operations are performed twice, and the results are added together to obtain 64 tensors with the same shape as the current feature. These tensors are then multiplied by the current feature and summed in the channel direction to obtain the cost volume. As \textit{grid sampling} is executed in software, it is reasonable to send the current feature from hardware to software, perform the rest of the processes in software, and pass the output to hardware. Compared to the case where the rest of the processes is performed in hardware, the communication between hardware and software can be reduced to 2/64. Furthermore, because the number of multiplications in CVF accounts for only 5.0$\%$ of the total, its execution latency can be hidden by parallelizing it with hardware execution as much as possible, even if its computation is slightly slow. \\
\textit{Operations not included in \cref{fig:muls}}: These are implemented in software for simplicity because a few calculations are required.
\fi
\todo{形式を大幅に修正}
\todo{日本語直す}

\subsection{HW Design}
\ifja
ここでは、高位合成ツールを用いて、HW 実装に適した処理を PL に実装する方法について説明する。まず、BN folding、量子化、近似を行って、アルゴリズムやパラメータを HW に適した形に変更する。そして、そのデザインの実装を行い、並列化を行う。
\else
Here, we describe the implementation of the hardware-friendly processes on PL using an HLS tool. First, the algorithms and parameters should be changed to make them suitable for hardware through BN folding, quantization, and approximation. The design is then implemented in hardware and parallelized.
\fi

\subsubsection{BN Folding}
\ifja
DeepVideoMVS のような近年の DNN モデルでは conv 層の後に BN 層を用いることが多い。BN は学習において、活性値を正規化することで精度を上昇させるために重要な役割を果たしているが、推論においては学習された固定のパラメータを用いてアフィン変換を行うだけである。そのため、計算回数を減らすために、推論では conv と結合して演算を行う \cite{qt}。
\else
Recent DNN models, including DeepVideoMVS, often use a BN layer after a convolution layer. The BN plays an important role in training to increase accuracy by normalizing activation values; however, in inference, it simply affine transforms them with fixed trained parameters. Therefore, in inference, the BN can be folded with the convolution to reduce the number of calculations \cite{qt}.
\fi


\subsubsection{Quantization}
\ifja
ここでは、PTQ (Post-Training Quantization) を用いて、学習済みのパラメータの量子化のみを行い、その他に変更は加えない。量子化はチャネルごとではなく、テンソルごとに行う。
\else
Here, we quantize the trained parameters using the PTQ (Post-Training Quantization) method, without any other modifications. Quantization is performed per tensor and not per channel.
\fi

\ifja
本研究で行った PTQ について conv を例に説明する。$\bm{W}, b, s$ をそれぞれ重み、バイアス、スケールとし、$\bm{x}, \bm{y}$ をそれぞれ入出力の活性値とすると、元の conv の計算は以下のように表される。
\begin{align}
    \bm{y}_{i,j} = \brak{\sum_{s,t} (\bm{W}_{s,t} \cdot \bm{x}_{i+s,j+t}) + b} \cdot s
\end{align}
PTQ した後の計算は $m_1, m_2$ を中間出力とすると、以下のように表される。
\begin{align}
    \hat{m}_1 &= \sum_{s,t} (\bm{\hat{W}}_{s,t} \cdot \bm{\hat{x}}_{i+s,j+t}) + \hat{b} \\
    \hat{m}_2 &= \hat{m}_1 \cdot \hat{s} \\
    \bm{\hat{y}}_{i,j} &= \textrm{clip} (\textrm{rshift} (\hat{m}_2, r))
\end{align}
$\hat{\cdot}$ がついた変数は PTQ 後の値を表し、$r$ は $\hat{m}_2$ と $\bm{\hat{y}}_{i,j}$ が表す値の範囲を揃えるために $\hat{m}_2$ を右シフトする量を表す。rshift は $\hat{m}_2$ を $r$ ビット右シフトして、四捨五入する操作を表している。clip は入力が $\bm{\hat{y}}$ の量子化ビットの範囲に収まるように、範囲外の値を clip している。
\else
We explain the PTQ performed in this study using \textit{conv} as an example. The \textit{conv} calculation before the PTQ is shown as
\begin{align}
    \bm{y}_{i,j} = \brak{\sum_{s,t} (\bm{W}_{s,t} \cdot \bm{x}_{i+s,j+t}) + b} \cdot s,
\end{align}
where $\bm{W}$, $b$, and $s$ denote the weight, bias, and scale, respectively, and $\bm{x}$ and $\bm{y}$ denote the activation values of the input and output, respectively. The calculation after the PTQ is as follows:
\begin{align}
    \hat{m}_1 &= \sum_{s,t} (\bm{\hat{W}}_{s,t} \cdot \bm{\hat{x}}_{i+s,j+t}) + \hat{b} \\
    \hat{m}_2 &= \hat{m}_1 \cdot \hat{s} \\
    \bm{\hat{y}}_{i,j} &= \textrm{clip} (\textrm{rshift} (\hat{m}_2, r)),
\end{align}
where $m_1$ and $m_2$ denote the intermediate outputs. The variables with $\hat{\cdot}$ indicate the values after the PTQ, and $r$ expresses a right shift amount of $\hat{m}_2$ to adjust the numerical ranges of $\hat{m}_2$ and $\bm{\hat{y}}_{i,j}$ values. Function \textit{rshift} expresses the $r$-bit right shift and rounding of $\hat{m}_2$. Function \textit{clip} clips the input out of range such that the input falls within the range of the quantization bit of $\bm{\hat{y}}$.
\fi

\ifja
まず、重み、バイアス、スケール、活性値の量子化ビットを決定する。この値については実際に量子化と推論を試して、目標の精度を満たす最小の値に設定することが望ましい。次に各値の量子化を行い、量子化パラメータを決定する。量子化パラメータとはここでは量子化した重み、バイアス、スケールの値、及び重み、バイアス、スケール、及び活性値を量子化するための乗数のことを表す。重み、バイアス、スケールに関しては、全ての値が量子化ビットの範囲内に収まるような最大の 2 の冪乗倍を行う。活性値の量子化のためにはまずデータセットからいくつかの画像、もしくはデータセットの画像と同じ平均と分散を持つランダムな画像を用意する。そして、それらを用いて推論することによって各層の活性値を集める。最後に、集めた活性値に対して、$\alpha \%$ 以上の値が量子化ビットの範囲内に収まるような最大の 2 の冪乗倍を行う ($\alpha$ は適当なクリッピング率)。すべての乗数を 2 の冪乗にすることで、いくつかの処理が簡略化できる。例えば、二つの活性値を足したり、結合したりする際には、両方の値の範囲が同じである必要がある。この PTQ の設定では、これらの範囲を揃えるために最大でも 1 つのシフト演算だけで十分になる。その結果、この設定を用いることで実装も簡潔になり、誤差も生まれにくく、割り算を使用する必要もなくなる。一方で、2 の冪乗以外の定数の乗数も許容した場合に比べて、ビット幅を有効に使いきれないという欠点はある。
\else
First, we determine the quantization bits for the weights, biases, scales, and activation values. These values should be selected as the smallest possible values that satisfy the target accuracy by attempting quantization and inference. Next, we quantize each value and determine the quantization parameters. The quantization parameters indicate the quantized values of the weights, biases, and scales, as well as the multipliers used to quantize the weights, biases, scales, and activation values. The weights, biases, and scales are multiplied by the largest power of two such that all values fall within the range of each quantization bit. To quantize the activation values, we first prepare several images from the dataset or random images with the same mean and variance as the dataset images. We then collect the activation values for each layer by inferring them. Finally, the collected values are multiplied by the largest power of two such that more than $\alpha\%$ values fall within the range of each quantization bit, where $\alpha$ is an appropriate clipping rate. Setting all multipliers to the powers of two simplifies several processes. For example, when two activation values are added or concatenated, their numerical ranges must be the same. In this PTQ setting, at most one left shift (\textit{lshift}) is sufficient to adjust these ranges. Consequently, this setting simplifies the implementation, reduces errors, and eliminates division operations. By contrast, the bit width cannot be used as effectively if constant multipliers other than those of powers of two are allowed.
\fi
\todo{Next to Subsequently}

\ifja
このように量子化することで、conv、addition、multiplication などの演算が整数の範囲内で計算できるようになり、HW 資源の節約とこれらの演算の高速化が期待できる。
\else
This quantization method allows operations, including \textit{conv}, \textit{addition}, and \textit{multiplication}, to be calculated within an integer range, which is expected to save hardware resources and accelerate these operations.
\fi

\subsubsection{LUT-based Approximation}
\ifja
本研究で使用した sigmoid や ELU の活性化関数は以下のように表される通り、exponential の演算が含まれている。
\begin{align}
    \textrm{sigmoid}(x) &= \dfrac{1}{1 + e^{-x}} \\
    \textrm{elu}(x) &=
    \left\{
        \begin{array}{ll}
            x & (x \geq 0) \\
            e^{x} - 1 & (x < 0)
        \end{array}
    \right.
\end{align}
これらを HW でそのまま実装すると実行時間が大きくなってしまう。そのため、LUT を利用して近似する。活性化関数の入力 $x$ の範囲 (例えば $-t \leq x \leq t$ など) をテーブルのエントリサイズで均等に分け、各エントリごとの結果の値をテーブルに格納しておく。ただし、指定した入力の範囲を超える入力が与えられた場合には LUT はテーブルの最も近い端の値を結果として返すこととする。この時、対称性を利用することで、sigmoid のテーブルサイズを半分に削減することができる。
\else
The \textit{sigmoid} and \textit{ELU} activation functions used in this study include exponential operations as follows:
\begin{align}
    \textrm{sigmoid}(x) &= \dfrac{1}{1 + e^{-x}} \\
    \textrm{elu}(x) &=
    \left\{
        \begin{array}{ll}
            x & (x \geq 0) \\
            e^{x} - 1 & (x < 0)
        \end{array}
    \right.
\end{align}
If these are straightforwardly implemented in hardware, the execution time will be large. Thus, we approximate them using LUTs. The range of their input $x$ (e.g., $-t \leq x \leq t$) is evenly divided by the number of table entries and the output values for each entry are stored in the table. However, if an input exceeding the specified input range is provided, the LUT returns the value at the closest end of the table. Here, the \textit{sigmoid} table size can be reduced by half by using symmetry.
\fi

\subsubsection{Overall Hardware Architecture}
\begin{figure}[tbp]
    \centering
    \includegraphics[width=\linewidth,pagebox=mediabox]{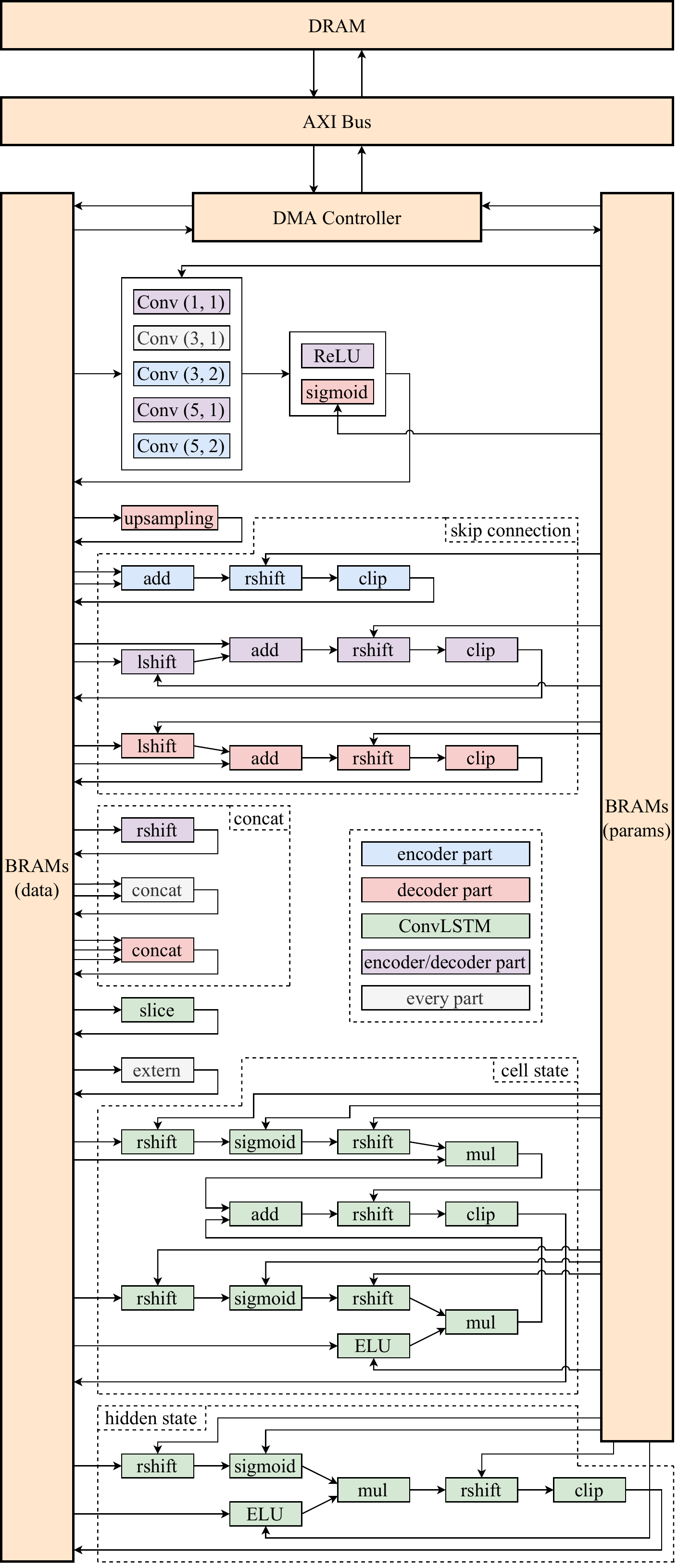}
    \ifja
    \caption{専用の算術演算パイプラインを含む全体的な HW アクセラレータのアーキテクチャ。extern は SW とのやり取りを行う演算を表す (詳しくは \cref{sec:extern} で後述)。}
    \else
    \caption{Overall hardware accelerator architecture including dedicated arithmetic pipelines. \textit{Extern} shows an operation to communicate with software (see \cref{sec:extern} for further details).}
    \fi
    \label{fig:hw-design}
\end{figure}


\ifja
次に、\cref{fig:hw-design} は PL 上に実装する処理の全体的な HW アクセラレータのアーキテクチャを示している。この回路内 (ただし、\cref{fig:hw-design} には示されていない) のハード・ワイヤードの FSM (Finite State Machine) が演算を自動で制御する。それぞれのステージでは種別ごとに演算器からなる専用の算術演算パイプラインが用意される。つまり、他のステージと同じ演算を行うステージについては HW 資源を再利用することができる。extern, shift, clip は \cref{tb:operations} には含まれていない演算だが、これらは活性値の量子化のために用いられている。また、addition, multiplication, shift などの element-wise operator は一つの値に対して連続的に処理を行えるため、これらの複数の演算は 1 つにまとめることができる。
\else
Next, we describe the overall hardware accelerator architecture of the processes to be implemented on the PL as shown in \cref{fig:hw-design}. A hard-wired finite state machine (FSM) in this circuit (not presented in \cref{fig:hw-design}) automatically controls the operations. Dedicated arithmetic pipelines consisting of operators are prepared for each stage according to the stage type. In other words, the circuits for an arithmetic pipeline or combination of operations in a stage can be reused if another stage performs the same pipeline and combination as those in this stage. \textit{Extern}, \textit{shift}, and \textit{clip}, which are not included in \cref{tb:operations}, are used to quantize the activation values. Additionally, because element-wise operators, such as \textit{addition}, \textit{multiplication}, and \textit{shift}, can continuously process a single value, the sequence of these operators can be folded into one.
\fi


\subsubsection{Data-level Parallelization}
\ifja
これまで述べたアーキテクチャに対して、データレベル並列化を施すことで、高速化を実現する。並列化の際には、実行速度と HW 資源や消費電力などの制約とのトレードオフを考慮することが重要である。本研究では HW 資源制約を満たす範囲で並列化を行う。conv が最も演算回数が多く、使用される回数も多いので、優先的に並列化する。conv は入力チャネル、出力チャネルごとに並列化ができ、理論上それらの並列度をかけた値倍だけ高速化される。その他の operator についても concatenation, slice, extern 以外はチャネルごとに並列化可能であり、理論上並列度倍の高速化が可能となる。
\else
We apply data-level parallelization to the architecture described thus far for further acceleration. When parallelizing operations, it is important to consider the trade-off between execution speed and practical constraints, such as hardware resources and power consumption. In this study, parallelization was performed such that hardware resource constraints were satisfied. As \textit{conv} has the largest number of multiplications, it should be preferentially parallelized. It can be parallelized for each input and output channel. Theoretically, the speed can be increased by a factor of the value obtained by multiplying the degrees of parallelism. Other operators, except for \textit{concatenation}, \textit{slice}, and \textit{extern}, can also be parallelized for each channel. Theoretically, the speed can be increased by a factor of the degree of parallelism.
\fi
\todo{文章がちょっとくどい}

\subsection{SW Design}

\ifja
SW 実装に適した処理について PL からデータを受け取って PL に指定された時に CPU で実行する方法を説明する。これらの処理を高速に実行できるように、これらの処理を通常の SW 開発と同じ手順・内容で、適切な言語を用いて実装する。
\else
We propose a design for the software-friendly processes, which can receive data from the PL and execute the process on the CPU when the PL specifies the process. To execute these processes at high speed, they should be implemented in an appropriate language using the same procedures and content as those for normal software development.
\fi
\todo{前半部を修正}

\ifja
SW 処理の高速化については様々な手法があるが、本研究では以下のような手法を用いている。
\begin{itemize}
    \item キャッシュヒット率を上げるようにメモリのアクセスパターンを最適化する
    \item 事前に確定している変数を埋め込む
    \item 量子化を行う
    \item マルチスレッド型の並列化を行う
\end{itemize}
使用している SW や言語の仕様に応じて、様々な最適化手法が考えられるが、一般に以上のような最適化手法を用いることで、SW での処理の高速化が可能である。
\else
Although there are various acceleration methods for software processing, the methods used in this study are as follows:
\begin{itemize}
    \item Optimize memory access patterns to increase the cache hit rate.
    \item Embed pre-determined variables.
    \item Perform quantization.
    \item Perform multithreaded parallelization.
\end{itemize}
Various optimization methods can be considered depending on the software and language specifications; however, in general, the above optimization methods can be used to speed up software processing.
\fi

\subsection{HW/SW Scheduling}
\ifja
最後に、HW 実装と SW 実装を組み合わせて、PL と CPU の処理の並列化も行い高速化する。この時、連続処理であることを活かした実行レイテンシーの隠蔽や、パイプラインにおいて律速になっている実行速度が遅い処理について高速化を行う。
\else
Finally, the hardware and software implementations are combined, and the PL and CPU processes are parallelized for acceleration. At this time, we hide the execution latencies by taking advantage of the fact that frames are given consecutively, and accelerate the time-consuming processes that are the bottlenecks in the pipeline.
\fi

\subsubsection{HW/SW Communication} \label{sec:extern}
\begin{figure}[tbp]
    \centering
    \includegraphics[width=0.8\linewidth,pagebox=mediabox]{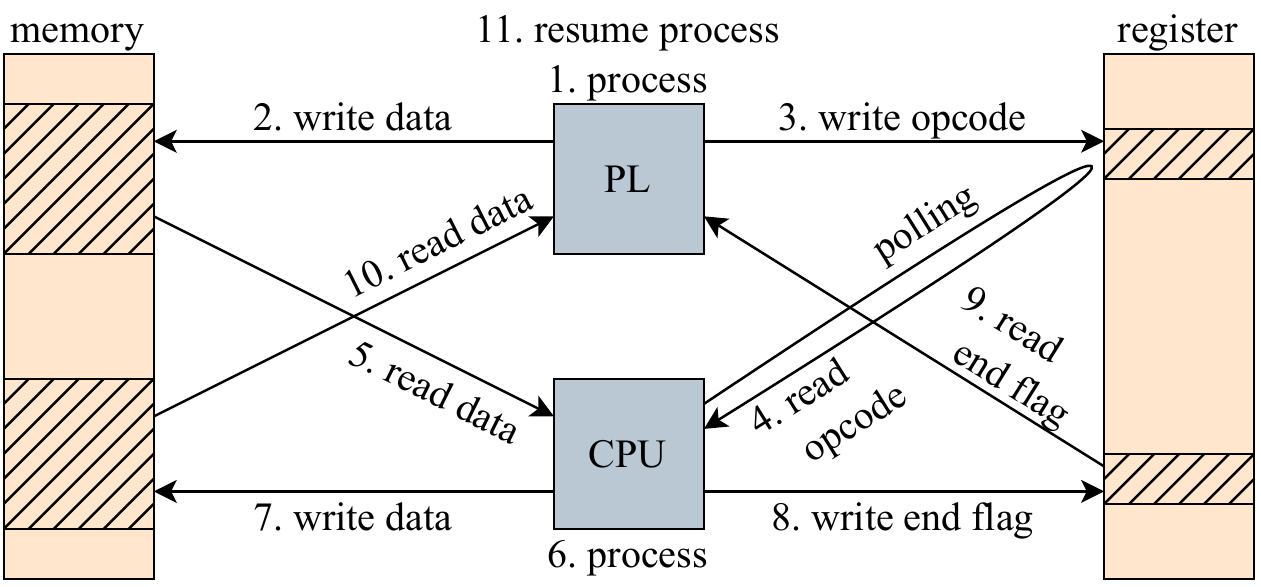}
    \ifja
    \caption{PL と CPU の間の割り込み処理機構の説明。この処理は \cref{fig:hw-design} の extern に対応する。}
    \else
    \caption{Illustration of the interrupt handling mechanism between the PL and CPU. This operation corresponds to \textit{extern} in \cref{fig:hw-design}.}
    \fi
    \label{fig:extern-operation}
\end{figure}

\ifja
それぞれの処理の終了通知やデータのやり取りを行うためには、HW・SW 間で通信を行う必要がある。この処理は \cref{fig:hw-design} の extern に対応しており、これは CMA (Contiguous Memory Allocator) と割り込み処理機構を用いて実現される。HW・SW 間の通信のためには、メモリ領域を共有する仕組みが必要となる。しかし、SW では仮想メモリ空間が扱えるのに対して、HW では物理メモリ空間しか扱えない。そのため、CMA を用いて確保した連続した物理メモリ領域を使用する。実行時にはそのメモリ領域の中でデータのやり取りを行うので、確保する領域は PL と CPU で共有されるデータを書き込むのに十分なサイズにする。通信のためには、\cref{fig:extern-operation} に示されている割り込み処理機構も用いる。具体的には、PL での処理が終わったら、メモリにデータを書き込んで、特定のレジスタに次に実行してほしい SW 処理を表す opcode を書き込む。それを CPU が polling で確認すると、メモリからデータを読み込んで、指定された処理を実行する。処理が終了すれば、メモリに処理結果を書き込み、特定のレジスタに終了通知フラグを立てる。そして、このフラグを HW が読み取って、続きの処理を再開するという流れである。
\else
To notify the end of each process and exchange data, communication between hardware and software is necessary. This operation corresponds to \textit{extern} in \cref{fig:hw-design}, which is achieved using a contiguous memory allocator (CMA) and interrupt handling mechanism. For communication between hardware and software, a mechanism to share memory space is required. However, hardware can only handle physical memory space, whereas software can handle virtual memory space. Thus, we use a contiguous physical memory area allocated using CMA. As data are exchanged within the memory area during execution, the allocated area should be sufficiently large to write the data shared by the PL and CPU. We also use the interrupt handling mechanism shown in \cref{fig:extern-operation} for the communication. Specifically, when the PL process finishes, the data are written to the memory, and an opcode representing the next CPU process is written to a specific register. When the CPU detects this by polling, it reads the data from the memory and executes the specified process. When the process is completed, the output is written to the memory and an end flag is set in a specific register. The PL then reads this flag and resumes the next process.
\fi
\todo{大幅に修正}

\subsubsection{Task-level Parallelization}
\begin{figure}[tbp]
    \centering
    \includegraphics[width=\linewidth,pagebox=mediabox]{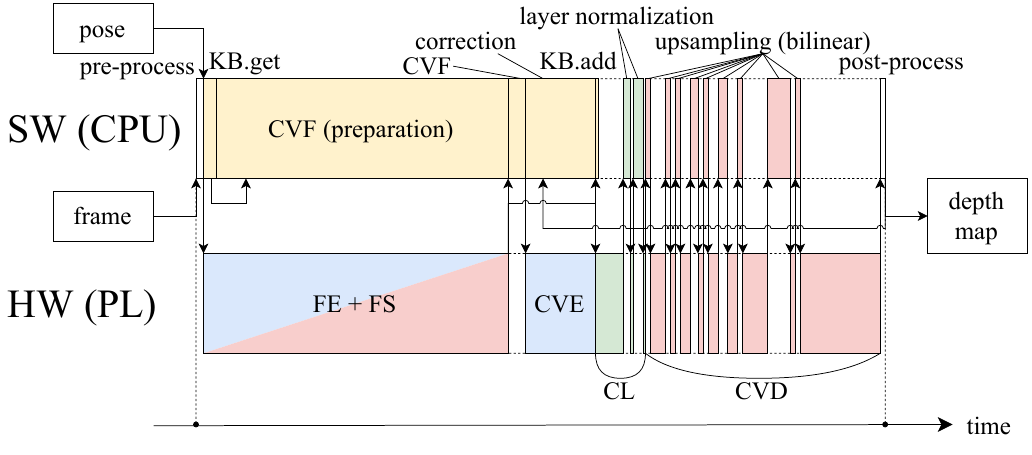}
    \ifja
    \caption{提案手法のパイプラインチャート。矢印はデータの依存関係を表している。各処理の長さは実際の実行レイテンシーに概ね対応している。}
    \else
    \caption{Pipeline chart of the proposed accelerator. Arrows indicate data dependencies. The length of each process roughly corresponds to the actual execution latencies.}
    \fi
    \label{fig:co-design}
\end{figure}

\ifja
\cref{fig:co-design} は完全に最適化されたパイプラインチャートを表している。最適化に至る前に HW と SW のそれぞれの処理の実行レイテンシーをプロファイリングする必要がある。そして、並列度を高め、実行レイテンシーを可能な限り隠蔽するには、各処理をどこに配置するのが最適かを検討することが重要である。特定の処理をさらに高速化するために工夫を凝らす必要があるかもしれない。
\else
\cref{fig:co-design} shows a fully optimized pipeline chart. Before optimization, it is necessary to profile the execution latencies of each process in hardware and software. It is also important to consider the best place for each process to increase parallelism and hide execution latencies as much as possible. It may be necessary to devise methods to further accelerate certain processes.
\fi

\ifja
例えば、本研究では、CVF は SW 処理の中で実行レイテンシーが最も長い処理で、このレイテンシーをいかに隠蔽できるかが高速化の鍵になっている。データの流れを詳しく解析すると、FS の出力が必要となる処理は一部分しかなく、grid sampling を含むそれ以外の処理 (CVF (preparation)) は HW が FE と FS を行っている間に並列して行うことができるとわかった。これにより、CVF にかかる全体のレイテンシーのうち、93$\%$ を隠蔽することができる。また、hidden state の補正に関しても CVE の処理と並列して行うことで、レイテンシーを隠蔽することができる。しかし、補正した hidden state が必要となる CL の処理の開始時には確実に補正処理が終わっていることを保証する必要がある。そこで、その時点で SW に割り込みを入れて、補正処理が終わるとすぐに HW 処理を再開するという手順をとっている。
\else
For example, in this study, CVF is the most time-consuming software process, and the key to the entire acceleration is determining how to hide this latency. A detailed analysis of the dataflow reveals that only a small part of the process requires the FS output, and the other part (CVF (preparation)), including \textit{grid sampling}, can be performed in parallel with the FE and FS execution in hardware. Thus, we can hide 93$\%$ of the total latency required for CVF. In addition, hidden state correction can also be performed in parallel with CVE execution to hide the latency. However, it is necessary to ensure that the correction process is completed at the beginning of CL, which requires the corrected hidden state. Thus, software is interrupted at the time, and the hardware process is resumed immediately after the correction process finishes.
\fi


\section{Evaluation}
\ifja
ここでは、提案手法を実際に実装し、性能を評価する。入出力の画像のサイズは $96 \times 64$ になるように設定し、DeepVideoMVS の著者が公開している、TUM RGB-D \cite{tum} を用いて事前学習されたモデルを使用して推論を行う。実装のためには Python を使用し、NNgen \cite{pyverilog,nngen} v1.3.3 と呼ばれるオープンソースの高位合成ツールを用いて、Verilog HDL のコードを生成する。そして、Vivado 2021.2 を用いて the ZCU104 board with Zynq UltraScale+ MPSoC XCZU7EV-2FFVC1156 from Xilinx 向けにビットストリームを生成する。最後に、PYNQ v2.6 for the ZCU104 board を用いて、提案手法を FPGA 上で実行する。PYNQ は Python API を持つ Jupyter を元にしたフレームワークを提供している。そこで、高速化が必要な SW 処理に関しては Cython v0.29 を用いて記述してコンパイルした後に、Python で実行する。
\else
In this section, we describe the implementation and performance evaluation of the proposed accelerator. The size of the input and output images is $96 \times 64$, and a model pre-trained with TUM RGB-D \cite{tum}, published by the authors of DeepVideoMVS, is used for inference. We use Python for the implementation and generate Verilog HDL codes using an open-source HLS tool called NNgen \cite{pyverilog,nngen} v1.3.3. We then generate a bitstream for the ZCU104 board with Zynq UltraScale+ MPSoC XCZU7EV-2FFVC1156 from Xilinx using Vivado 2021.2. Finally, we execute the proposed accelerator on the FPGA using PYNQ v2.6 for the ZCU104 board. PYNQ provides a Jupyter-based framework with Python APIs. Thus, the software processes that must be accelerated are written and compiled using Cython v0.29 to be executed in Python.
\fi

\ifja
実装におけるパラメータについては以下のように設定する。重み、バイアス、スケール、活性値の量子化ビットはそれぞれ 8, 32, 8, 16 である。活性値の量子化においては、$\alpha = 95 \%$ に設定する。また、LUT 基づく近似においては、エントリ数を 256、入力の範囲 $t$ を 8 に設定する。並列度については、conv の入力チャネル方向は 2、出力チャネル方向は kernel size が 5 の時のみ 2、それ以外の場合は 4 に設定する。その他の並列化可能な operator については、並列度をチャネル方向に 4 としている。SW 処理に関しては Xilinx ZCU104 board が 2 コアのため、並列度を 2 にしている。
\else
In the implementation, we set the parameters as follows. The quantization bits for the weights, biases, scales, and activation values are 8, 32, 8, and 16, respectively. To quantize the activation values, we set the clipping rate $\alpha$ to 95$\%$. In LUT-based approximation, we set the number of entries to 256 and the input range $t$ to 8.0. The degree of parallelism for \textit{conv} was set to 2 in the input channel direction, 2 only when the kernel size was 5, and 4 in the other cases in the output channel direction. For other parallelizable operators, the degree was set to 4 in the channel direction. For the software processes, the degree was set to 2 because the Xilinx ZCU104 board has two cores.
\fi

\ifja
7-Scenes \cite{7-scenes} というデータセットを使用し、その中の 8 個のシーン (chess/seq-01, chess/seq-02, fire/seq-01, fire/seq-02, office/seq-01, office/seq-03, redkitchen/seq-01, redkitchen/seq-07) を用いて提案手法の評価を行う。誤差については出力と ground truth の間の MSE (Mean Squared Error) を用いて計算を行っているので、低いほど良い結果を示していることになる。
\else
We used 7-Scenes dataset \cite{7-scenes} to evaluate the proposed accelerator using the following eight scenes: chess/seq-01, chess/seq-02, fire/seq-01, fire/seq-02, office/seq-01, office/seq-03, redkitchen/seq-01, and redkitchen/seq-07. As the error is calculated using the mean squared error (MSE) between the output and ground truth, a lower error value indicates a better output.
\fi

\subsection{Execution Time}
\begin{table}[tbp]
    \ifja
    \caption{PL と CPU を用いた提案手法の実行と CPU のみでの C++ 実装の処理の実行時間の比較。}
    \else
    \caption{Comparison of the execution time per frame between the processing of the proposed accelerator on the PL and CPU and that of the C++ implementation on the CPU-only.\todo{std なくてもいい？}}
    \fi
    \label{tb:speed}
    \centering
    \begin{tabular}{c|rrr}
        \hline \hline
        Platform          &     median [s] & std [s] &  frequency [MHz] \\
        \hline
        CPU-only          &        16.744  &   0.049 &             N/A  \\
        CPU-only (w/ PTQ) &        13.248  &   0.035 &             N/A  \\
        PL + CPU (ours)   & \textbf{0.278} &   0.118 & \textbf{187.512} \\
        \hline
    \end{tabular}
\end{table}

\ifja
PL と CPU を用いた提案手法の実行と CPU のみでの C++ 実装の処理の実行時間を Xilinx ZCU104 board を用いて比較する。C++ で実装したものはボード上で g++ 7.3.0 を用いて -O3 オプションをつけてコンパイルする。提案手法では、クロック周波数は 187.512 MHz で、これは Vivado のタイミング制約を満たすような最大のクロック周波数となっている。\cref{tb:speed} は評価用のデータセットの一枚の画像の処理が終わるまでの時間の中央値と標準偏差を示している。C++ での実装を CPU のみで実行した場合は、PTQ を行うことで若干の処理時間の減少は見られるが、それほど大きなものではない。提案手法では、大幅に実行時間が短くなり、CPU のみで実行した場合と比べて 60.2 倍の高速化となっている。
\else
We compared the execution time between the processing of the proposed accelerator on the PL and CPU and that of the C++ implementation on the CPU-only using the Xilinx ZCU104 board. The C++ implementation was compiled using g++ 7.3.0 with the -O3 option on the board. The clock frequency of the proposed accelerator was 187.512 MHz, which is the frequency at which all timing constraints are met in Vivado. \cref{tb:speed} shows the median and standard deviation of the time to finish processing a single image in the evaluation datasets. When the C++ implementation was run on the CPU-only, PTQ led to a slight reduction in execution time but not to a significant extent. The proposed accelerator drastically reduced the time, completing the tasks 60.2 times faster than the CPU-only execution.
\fi

\ifja
HW・SW 間の通信に用いる extern の処理にかかる overhead についても考える。overhead とは、HW 処理が割り込みを入れてからすぐに SW 処理が開始できる状況の時に、HW での待ち時間と SW での処理時間の差のことと定義する。overhead には SW がデータを読み書きする時間やその他の制御にかかる時間が含まれている。この overheads を計測すると、中央値で 4.7 ms となっており、これは全体の実行時間の 1.69$\%$ である。そのため、HW と SW を協調させるメリットの方が overhead よりも大きい。
\else
We also considered the overhead required to perform \textit{extern}, which is used to communicate between hardware and software. Overhead is defined as the difference between the waiting time in hardware and processing time in software, assuming that software processing can start immediately after being interrupted by hardware. The overhead includes the time required for software to read/write data and other control times. The median value of the measured overhead is 4.7 ms, which is 1.69$\%$ of the total execution time. Thus, the advantage of coordinating hardware and software outweighs the overhead.
\fi

\subsection{Hardware Resources}
\begin{table}[tbp]
    \ifja
    \caption{提案手法の Xilinx ZCU104 board における HW 資源の使用状況。}
    \else
    \caption{Hardware resource utilization of the proposed accelerator on the Xilinx ZCU104 board.}
    \fi
    \label{tb:resources}
    \centering
    \begin{tabular}{c|rrr}
        \hline \hline
        Name  & \#Utilization & Available & Utilization [$\%$] \\
        \hline
        Slice &         28256 &     28800 &      \textbf{98.1} \\
        LUT   &        176377 &    230400 &              76.6  \\
        FF    &        143072 &    460800 &              31.0  \\
        DSP   &           128 &      1728 &               7.41 \\
        BRAM  &           309 &       312 &      \textbf{99.0} \\
        \hline
    \end{tabular}
\end{table}

\ifja
提案手法の Xilinx ZCU104 board における HW 資源の使用状況は \cref{tb:resources} の通りである。slice や BRAM はほぼ使い切っており、HW 資源を最大限有効活用できていることがわかる。
\else
\cref{tb:resources} shows the hardware resource utilization of the proposed accelerator on the Xilinx ZCU104 board. The slices and block RAMs (BRAMs) are almost fully used, indicating that we can take full advantage of the hardware resources.
\fi

\subsection{Accuracy}
\newcommand{\imgadir}{fire-seq-01}
\newcommand{\imga}{000139}
\begin{figure*}[t!]
    \centering
    \begin{minipage}[t]{0.15\linewidth}
        \centering
        \includegraphics[keepaspectratio,width=\linewidth,pagebox=mediabox]{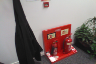}
        \subcaption{Input}
    \end{minipage}
    \begin{minipage}[t]{0.15\linewidth}
        \centering
        \includegraphics[keepaspectratio,width=\linewidth,pagebox=mediabox]{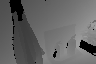}
        \subcaption{Ground truth}
    \end{minipage}
    \begin{minipage}[t]{0.15\linewidth}
        \centering
        \includegraphics[keepaspectratio,width=\linewidth,pagebox=mediabox]{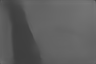}
        \subcaption{Output of C++ impl}
    \end{minipage}
    \begin{minipage}[t]{0.15\linewidth}
        \centering
        \includegraphics[keepaspectratio,width=\linewidth,pagebox=mediabox]{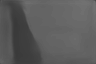}
        \subcaption{Output of C++ impl w/ PTQ}
    \end{minipage}
    \begin{minipage}[t]{0.15\linewidth}
        \centering
        \includegraphics[keepaspectratio,width=\linewidth,pagebox=mediabox]{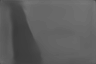}
        \subcaption{Output of the proposed accelerator}
    \end{minipage}
    \ifja
    \caption{シーン {\imgadir}、フレーム番号 {\imga} の処理の結果。ground truth との間の MSE はそれぞれ (c) 0.091, (d) 0.073, (e) 0.089, (f) 0.084 である。}
    \else
    \caption{Results of processing the frame number {\imga} in the {\imgadir} scene. The MSEs between the outputs and ground truth are (c) 0.091, (d) 0.073, (e) 0.089, and (f) 0.084, respectively.}
    \fi
    \label{fig:frame-a}
\end{figure*}

\newcommand{\imgcdir}{redkitchen-seq-07}
\newcommand{\imgc}{000268}
\begin{figure*}[t!]
    \centering
    \begin{minipage}[t]{0.15\linewidth}
        \centering
        \includegraphics[keepaspectratio,width=\linewidth,pagebox=mediabox]{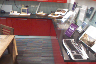}
        \subcaption{Input}
    \end{minipage}
    \begin{minipage}[t]{0.15\linewidth}
        \centering
        \includegraphics[keepaspectratio,width=\linewidth,pagebox=mediabox]{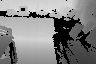}
        \subcaption{Ground truth}
    \end{minipage}
    \begin{minipage}[t]{0.15\linewidth}
        \centering
        \includegraphics[keepaspectratio,width=\linewidth,pagebox=mediabox]{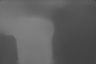}
        \subcaption{Output of C++ impl}
    \end{minipage}
    \begin{minipage}[t]{0.15\linewidth}
        \centering
        \includegraphics[keepaspectratio,width=\linewidth,pagebox=mediabox]{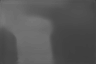}
        \subcaption{Output of C++ impl w/ PTQ}
    \end{minipage}
    \begin{minipage}[t]{0.15\linewidth}
        \centering
        \includegraphics[keepaspectratio,width=\linewidth,pagebox=mediabox]{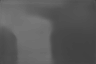}
        \subcaption{Output of the proposed accelerator}
    \end{minipage}
    \ifja
    \caption{シーン {\imgcdir}、フレーム番号 {\imgc} の処理の結果。ground truth との間の MSE はそれぞれ (c) 0.808, (d) 0.880, (e) 1.099, (f) 1.050 である。}
    \else
    \caption{Results of processing the frame number {\imgc} in the {\imgcdir} scene. The MSEs between the outputs and ground truth are (c) 0.808, (d) 0.880, (e) 1.099, and (f) 1.050, respectively.}
    \fi
    \label{fig:frame-c}
\end{figure*}

\begin{figure}[tbp]
    \centering
    \includegraphics[width=0.8\linewidth,pagebox=mediabox]{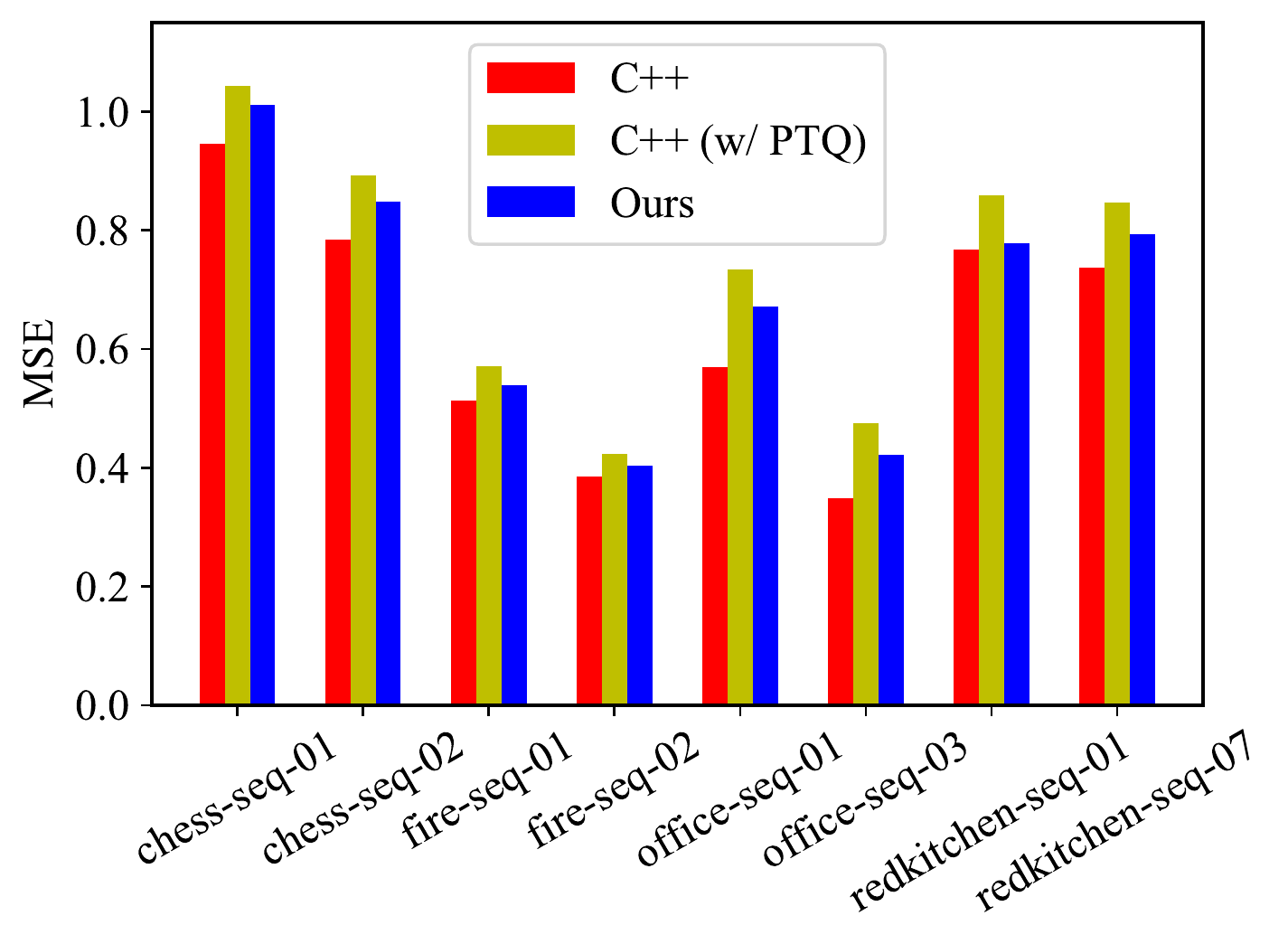}
    \ifja
    \caption{提案手法から求めた MSE と C++ 実装から求めた MSE のシーンごとの差異。}
    \else
    \caption{Scene-by-scene difference between MSE obtained from the proposed accelerator and MSE obtained from the C++ implementation.}
    \fi
    \label{fig:errors}
\end{figure}

\ifja
PTQ や LUT に基づく近似を行ったことによる精度の低下がどの程度であるかを評価する。まず、\cref{fig:frame-a,fig:frame-c} を用いて定性的な評価を行う。提案手法の出力は C++ の実装の出力と比べて、見た目で区別できるほどの大きな精度劣化はないことがわかる。次に \cref{fig:errors} を用いて定量的な評価を行う。どのデータセットにおいても C++ の実装と比べると若干精度は落ちているが、ほとんどの場合で精度劣化は 10$\%$ 未満に抑えられていることがわかる。また、C++ w/ PTQ の方が提案手法よりも精度が低いのは、提案手法では右シフトを行うときに round を行っているが、C++ w/ PTQ では行っていないことが原因だと考えられる。
\else
We evaluated the degree of accuracy loss due to PTQ and LUT-based approximation. First, a qualitative evaluation was performed using \cref{fig:frame-a,fig:frame-c}. The outputs of the proposed accelerator did not exhibit sufficient degradation in terms of accuracy to be visually distinguishable from the outputs of the C++ implementation. Next, we performed a quantitative evaluation using \cref{fig:errors}. Although the accuracy was slightly degraded compared with the C++ implementation for all scenes, the accuracy degradation remained below 10$\%$ in most cases. The accuracy of C++ with PTQ was lower than that of the proposed accelerator, which can be attributed to the fact that the proposed accelerator performs rounding after right shifts, whereas C++ with PTQ does not.
\fi

\section{Related Work}

\subsection{FPGA-based DNN Accelerator}
\ifja
DNN は近年さらに精度が向上しているが、その代償として複雑性が増している。一方で、このような高精度の DNN による推論をエッジや組み込みの小規模のデバイスで高速に低消費電力で行うことも求められている。そのようなデバイスとして特に注目されている FPGA を用いて、DNN を高速化するという研究は数多く行われている。ICAN \cite{ican} では、3D のデータタイリングを工夫することで、CNN の効率的な FPGA 実装を行っている。また、\cite{mnist-accel} では入力と重みの値を並列に読み込めるようにメモリの配置を工夫したり、二値化ニューラルネットである XNOR-Net \cite{xnor-net} を用いた計算回数や HW 資源の削減をおこなったりしている。DNNBuilder \cite{dnnbuilder} や DNNExplorer \cite{dnnexplorer} では、FPGA を用いた DNN accelerator の自動設計を試みている。自動設計には、メモリアクセスの効率化を行ったり、使用可能な HW 資源量と DNN の複雑性や特性を考慮した並列化の方式を求めたりする必要がある。このような手法の評価で使用されている DNN は最大でも 38 層であり、より最近のより大規模な DNN の FPGA 実装はあまり研究されていない。
\else
In recent years, DNNs have been further improved in terms of accuracy at the expense of increased complexity. By contrast, on small edge or embedded devices, it is necessary to perform inference on such high-accuracy DNNs at high speed and low power consumption. Several studies have been conducted on DNN acceleration using FPGAs because FPGAs have attracted particular attention for such devices. ICAN \cite{ican} efficiently implements CNNs on an FPGA by devising a 3D data tiling. In addition, \cite{mnist-accel} devises memory allocation to read input and weight values in parallel, and reduces the number of calculations and hardware resource usage using XNOR-Net \cite{xnor-net}, a binary neural network. DNNBuilder \cite{dnnbuilder} and DNNExplorer \cite{dnnexplorer} design FPGA-based DNN accelerators automatically. Automatic design requires improving the efficiency of memory access and seeking a parallelization scheme by considering the amount of available hardware resources and the complexity and characteristics of the DNN. These methods use DNNs with a maximum of 38 layers for evaluation, and few studies have been conducted on implementing more modern, larger DNNs on FPGAs.
\fi

\subsection{HW/SW Co-design}
\ifja
軽量かつ高速な実装を行いたいという背景から、HW/SW 協調設計を行う事例も増えている。特に動画像処理の一種である、VO (Visual Odometry) algorithm \cite{vo} や HOG (Histograms of Oriented Gradients) \cite{hog}、ADF (Anisotropic Diffusion Filter) \cite{fadf} などを FPGA を用いて高速化する手法が複数提案されている。これらは古典的なフィルタ処理だが、機械学習手法の一つである、Decision Forest \cite{df} や EfficientNet \cite{efnet} を高速化する手法についても提案されている。しかしながら、これらは単体のアルゴリズムの処理の高速化であり、動画像処理や機械学習を組み合わせたような大規模で実用的なアプリケーションにおける高速化手法についてはあまり研究がなされていない。
\else
HW/SW co-design has become increasingly popular for achieving lightweight and high-speed implementation. Several methods have been proposed to accelerate image/video processing, such as the visual odometry (VO) algorithm \cite{vo}, histograms of oriented gradients (HOG) \cite{hog}, and anisotropic diffusion filter (ADF) \cite{fadf}, using FPGAs. Although these methods are based on classical filtering methods, methods to accelerate the decision forest \cite{df} and EfficientNet \cite{efnet}, which are machine learning techniques, have also been proposed. However, these methods aim to accelerate stand-alone algorithms, and few studies have been conducted on acceleration methods for large-scale, practical applications, such as depth estimation that combines video processing and machine learning.
\fi

\section{Conclusion}
\ifja
本稿では、三次元空間の再構成に用いられる DNN ベースの深度推定手法の一つである DeepVideoMVS を用いた FPGA ベースの新しい高速化手法を提案する。DeepVideoMVS は、動画像処理特有の処理と DNN を組み合わせているため、全体の処理を最適化するのが難しい。そこで、その手法固有の特性に合わせて、最近の SoC FPGA 上の PL と CPU のような異種のコンポーネントを適切に利用するための HW/SW co-design を用いる。HW 実装するのに適していない演算もあるので、各演算の回数やメモリアクセスのパターンを分析し、HW 実装の容易さと HW によって期待される高速化の度合いという包括的な側面を考慮した上で、SW 実装を行う演算を決める。HW 実装と SW 実装は互いの実行レイテンシーを隠蔽するために、PL と CPU が並列に協調して動作するように実行される。提案したアクセラレータは Xilinx ZCU104 board 上に実装した。その結果、提案手法では、精度の低下を最小限に抑えて、SW のみの実装と比べて 60.2 倍の高速化を達成することができた。
\else
This paper proposed a novel FPGA-based accelerator for DeepVideoMVS, which is a DNN-based depth estimation method for 3D reconstruction. DeepVideoMVS combines traditional image/video processing algorithms and DNNs, making it difficult to optimize the entire process. Thus, we employed HW/SW co-design to appropriately utilize heterogeneous components in modern SoC FPGAs, such as PL and CPU, according to the inherent characteristics of the method. As some operations are unsuitable for hardware implementation, we determined the operations to be implemented in software through analyzing the number of times each operation is performed and its memory access pattern, and then considering comprehensive aspects: the ease of hardware implementation and degree of expected acceleration by hardware. The hardware and software implementations were executed in parallel on the PL and CPU to hide their execution latencies. The proposed accelerator was developed on a Xilinx ZCU104 board. Experiments showed that the proposed accelerator operates 60.2 times faster than the software-only implementation on the same FPGA board with minimal accuracy degradation.
\fi

\ifja
一方で、HW 資源の削減や速度・精度の改善についてはさらに検討の余地が残されている。例えば量子化手法や畳み込み層の HW 実装の手法については次々に新しい手法が提案されているため、そのような手法を提案手法に適用することで、さらに小規模、高速、高精度な実装が達成できるのではないかと考えられる。
\else
However, there is room for further reduction in hardware resource usage and improvement in terms of speed and accuracy. For example, because several new methods have been proposed for quantization and hardware implementation of convolution layers, smaller, faster, and more accurate implementations can be achieved by applying these techniques to the proposed accelerator.
\fi

\section*{Acknowledgment}
\addcontentsline{toc}{section}{Acknowledgment}
This work is supported in part by JSPS KAKENHI 19H04075 and 18H05288, JST CREST JPMJCR21D2, and the collaboration research with KONICA MINOLTA.



\end{document}